# The Alignment of Molecular Cloud Magnetic Fields with the Spiral Arms in M33

Hua-bai Li and Thomas Henning

*Max-Planck Institute for Astronomy, Königstuhl 17, D-69117 Heidelberg, Germany*

**The formation of molecular clouds, which serve as stellar nurseries in galaxies, is poorly understood. A class of cloud formation models suggests that a large-scale galactic magnetic field is irrelevant at the scale of individual clouds, because the turbulence and rotation of a cloud may randomize the orientation of its magnetic field[1,2]. Alternatively, galactic fields could be strong enough to impose their direction upon individual clouds[3,4], thereby regulating cloud accumulation and fragmentation[5], and affecting the rate and efficiency of star formation[6]. Our location in the disk of the Galaxy makes an assessment of the situation difficult. Here we report observations of the magnetic field orientation of six giant molecular cloud complexes in the nearby, almost face-on, galaxy M33. The fields are aligned with the spiral arms, suggesting that the large-scale field in M33 anchors the clouds.**

At a distance about 900 kpc[7], M33 is the nearest face-on galaxy with pronounced optical spiral arms. To resolve a typical giant molecular cloud (GMC) with a size of tens to a hundred parsecs, we used the Submillimeter Array[8] (SMA), which offers a linear spatial resolution of ~15 pc at 230 GHz (the frequency of the CO J = 2-1 transition) at the distance of M33 using the array's most compact configuration. In order to observe the strongest CO line emission, we picked the six most massive GMCs from M33[9]. It is clear to which spiral arms most GMCs are related, except for GMC3, which is situated in the middle of two optical arms (Fig. 1).

We determined the orientations of the GMC magnetic fields (B-fields) from the polarization of CO emission lines, which should be either perpendicular or parallel to the local B-field direction projected on the sky (the Goldreich–Kylafis effect[10]). Though there are other B-field tracers[11] that do not have this 90° ambiguity, CO is much more abundant and allows current radio telescopes to perform extragalactic cloud observations. Despite the 90° ambiguity, such a B-field observation is still valuable[12]. An intrinsically random field distribution, as occurs when the turbulence is super-Alfvenic[13] (i.e., when turbulent energy dominates B-field energy), will still be random with this ambiguity. On the other hand, an intrinsically single-peaked Gaussian-like field distribution, in the presence of sub-Alfvenic turbulence[13] (i.e., when B-field energy dominates turbulent energy), will either remain single-peaked, or split into two peaks approximately 90° apart ("double peaks"). From the total distribution of the offsets between the CO polarization of the M33 GMCs and the local arm directions (Fig. 2), the trend of double peaks is clearly visible (Fig. 3). The distribution can be fitted by a double-Gaussian function with peaks at -1.9º ± 4.7º and 91.1º ± 3.7º and a standard deviation of 20.7º ± 2.6º. This result is barely affected if the inter-arm GMC3 is excluded.

The angle dispersion ($\Sigma$) of the CO polarization and its offset ($\Delta$) from the arm directions are determined by the dispersion of B-field orientations ($\sigma$), the dispersion of the offsets between cloud mean fields and spiral arms ($\delta$), the observational errors ($\varepsilon < 10°$), and the Goldreich–

Kylafis effect. By performing Monte Carlo simulations (Supplementary Information), we can estimate the likelihood of observing $\Sigma$ and $\Delta$ within certain ranges, and determine which combinations of $\sigma$ and $\delta$ are able to produce the observed confidence level (Fig. 4). Random fields or field-arm offsets are very unlikely. Only when $\sigma$ = 17°–22° with $\delta$ < 8° do the simulations give a similar confidence level as the observed. This indicates that the mean field directions are well-defined and highly correlated with the spiral arms, which is consistent with the picture that the B-fields in the GMCs are compressed within the spiral arms, and the fields can exert tension forces (because of $\delta$, see Supplementary Fig. 1) strong enough to resist cloud rotation ("magnetic braking" [4, 9]). To accrete the mass to form a cloud from the accumulation length scale (hundreds of pc) in a shearing galactic disk, cloud rotation is inevitable due to conservation of angular momentum[1, 2], unless the momentum is consumed by other mechanism, e.g., magnetic tension. The fact that the GMCs in M33 show significantly smaller angular momenta than predicted by the Toomre instability criterion[9] agrees with our observations. Fig. 3 is also consistent with the cloud B-fields being perpendicular to the arms; however, there is neither a known mechanism nor a numerical simulation supporting such a field configuration.

The CO polarization is not necessarily aligned with the local synchrotron polarization (Fig. 3), which traces the B-fields in regions near the GMCs that are warmer and more diffuse[14] than the clouds themselves. From the viewpoint of the dynamo theory, it is possible for cloud fields to decouple from the large-scale galactic fields due to the small volume filling factor of molecular clouds[15]. This is consistent with the decoupling of synchrotron/CO polarization, but fails to explain the correlation between cloud B-fields and spiral arms. To our knowledge, no numerical simulations of small-scale dynamos within molecular clouds have been able to produce ordered field directions coherent with spiral arms. However, density wave compression, the cause of spiral arms, can explain the observation as follows. In the classical picture of inter stellar media[16], cold and warmer phases are in pressure equilibrium. While orbiting through the gravitational well of the spiral arms, the cold phase experiences a much stronger shock compression than the warmer phase does[17]. Therefore, it is unsurprising that the orientation of the B-fields in the warmer media (traced by synchrotron radiation) can decouple from the morphology of the more compressed fields in cold media (traced by CO). As the gas and B-fields in the cold media are compressed together, they should follow the same spiral morphology as long as the subsequent cloud formation activities do not disturb the field orientations significantly.

The non-random B-fields imply that the cloud turbulence is sub-Alfvenic. For this kind of turbulence, we can follow Chandrasekhar and Fermi[18], assuming the dispersion in B-field direction is coupled with the lateral velocity of gas turbulence, to estimate the B-field strength: $B_{pos} = 9.3\sqrt{n(H_2)}\frac{v}{\sigma}$, where $B_{pos}$, $n(H_2)$ and $v$ are, respectively, the plane-of-sky component of the B-field (micro-Gauss), $H_2$ density (cm$^{-3}$), and FWHM spectral linewidth (km/s)[19]. With $\sigma$ = 17°–22°, $n(H_2)$ = 10$^3$–10$^4$ cm$^{-3}$ [20], and $v$ ~ 10 km/s [9], the $B_{pos}$ is 0.1–0.5 milli-Gauss. This value is comparable to that of B-fields in molecular clouds of the Milky Way, and is about 100 times greater than the B-field strength estimated from synchrotron observations[21].

Evidence of sub-Alfvenic turbulence is also observed within clouds of the Milky Way. The Galactic B-field also anchors molecular clouds[11, 22], and aligns the velocity anisotropy of turbulence[5, 23]. From our edge-on view of the disk, the Milky Way fields, however, have rich structures at the hundred-parsec scale (the scale of the GMCs), instead of being aligned with the disk[11]. If galactic dynamics are similar for the Milky Way and M33, the simplest explanation is that B-fields of spiral arms can have much more structure perpendicular to the disk (as observed from an edge-on view) than within the disk plane (as observed from a bird's-eye view). Several mechanisms can help with this anisotropic B-field structures. Firstly, the density wave compression occurs mainly in a direction parallel to the disk plane, and secondly, the Parker instability[24] and stellar feedback (e.g., bipolar giant HII bubbles[25] and galactic fountains[26]) concentrate along gaseous arms, tending to deviate the B-fields toward directions perpendicular to the disk. A bird's-eye view of cloud B-fields is currently much more difficult to acquire compared to an edge-on view, but can offer important new insights into GMC/galaxy dynamics. Next-generation array telescopes (e.g. the Atacama Large Millimeter/submillimeter Array) will be more competent to efficiently survey molecular clouds from different views.

**Acknowledgements** We thank the referees, Erik Rosolowsky, Rahul Shetty, T. K. Sridharan, Martin Houde, Scott Paine, Hsiang-Hsu Wang, Alexander Karim, Sarah Ragan, Kester Smith, Paul Boley, and Tingting Wu for comments. We appreciate the helps from Dan Marrone, Glen Petitpas, and Ram Rao with the observations. We are grateful for the Herschel maps of M33 offerd by Carsten Kramer. This research is supported by Max-Planck-Institut für Astronomie and Harvard-Smithsonian Center for Astrophysics. The Submillimeter Array is a joint project between the Smithsonian Astrophysical Observatory and the Academia Sinica Institute of Astronomy and Astrophysics and is funded by the Smithsonian Institution and the Academia Sinica.

**Author Contributions** The experiment were designed and executed by H.L.. H.L. and T.H. contributed jointly to the manuscript.

**Author Information** Correspondence and requests for materials should be addressed to H.L. (li@mpia.de)

**Fig 1: The optical spiral arms and the locations of the six most massive GMCs in M33**. The background is an optical image of M33[21]. The vectors show the 3.6 cm synchrotron polarization[21], with the telescope beam size shown in the lower left corner. The vector length is proportional to the intensity of polarized emission. A sketch of the optical arms[27] is shown in light-purple solid lines. The contours show the structures between 36" – 48" derived from scale decomposition[28] of the 500 μm Herschel data[29] (the lowest contour level has 30% of the peak intensity, and the following levels increase linearly with 10% peak intensity). The GMC locations ("+"s) are numbered 1-6. The optical arms related to GMC 1, 5, and 6 are clear. GMC 3 is in the middle between two arms. GMC 2 and 4 are on the extensions (dark-purple dashed lines) from, respectively, two solid lines. GMC 4 just takes a short straight extension. For GMC 2, we adopt the southern arm defined by Rogstad, Wright & Lockhart[30] as the extension, which well traces the 500 μm clumps. The arms are traced slightly differently in various literatures, and this observational uncertainty will contribute to $\delta$, the dispersion of polarization-arm offset (Supplementary Information).

**Fig 2: CO (2-1) maps and polarization vectors.** The contours are 90%, 80%, … 10% of the peak intensity of each cloud. The red vectors show polarization detections, for which the ratio of the polarization level to its uncertainty is greater than 3 and the error in direction is less than 10°. The thick gray vectors show the tangents of the local optical arms[27, 30] for GMC 1, 2, 4, 5, and 6. For the inter-arm GMC 3, the gray vector shows the mean of the two tangents of the nearby arms at the positions closest to GMC 3. The ellipses indicate the SMA synthetic beams; all detections are spatially independent. The coordinate (R.A., Dec) offsets are in arc-seconds.

**Fig. 3: Distribution of the CO polarization-arm offsets.** The offsets are from the difference between the red and gray vectors in Fig. 2. Contributions from different GMCs are distinguished by the colors. The distribution can be fitted by a double-Gaussian function with a standard deviation of 20.7° ± 2.6° and peaks at -1.9° ± 4.7° and 91.1° ± 3.7°. The directions of synchrotron polarization from the regions near the GMCs (within one beam size, which is shown in Fig. 1) are also shown as the dashed lines, with the same color code as the GMCs.

**Fig. 4: Likelihood of obtaining simulated angle dispersions ($\Sigma$) and offsets ($\Delta$) of CO polarization within the observed 90% confidence intervals.** The 90% confidence intervals from the data in Fig. 3 are 15.1° < $\Sigma$ < 26.4 and $|\Delta|$ < 10.7°. The likelihood is estimated by Monte Carlo simulations (Supplementary Information) with various combinations of $\sigma$ (B-field dispersion) and $\delta$ (dispersion of the offsets between mean fields and arms). Only $\sigma \sim 20°$ with $\delta$ < 8° can give a similar confidence level within the intervals.

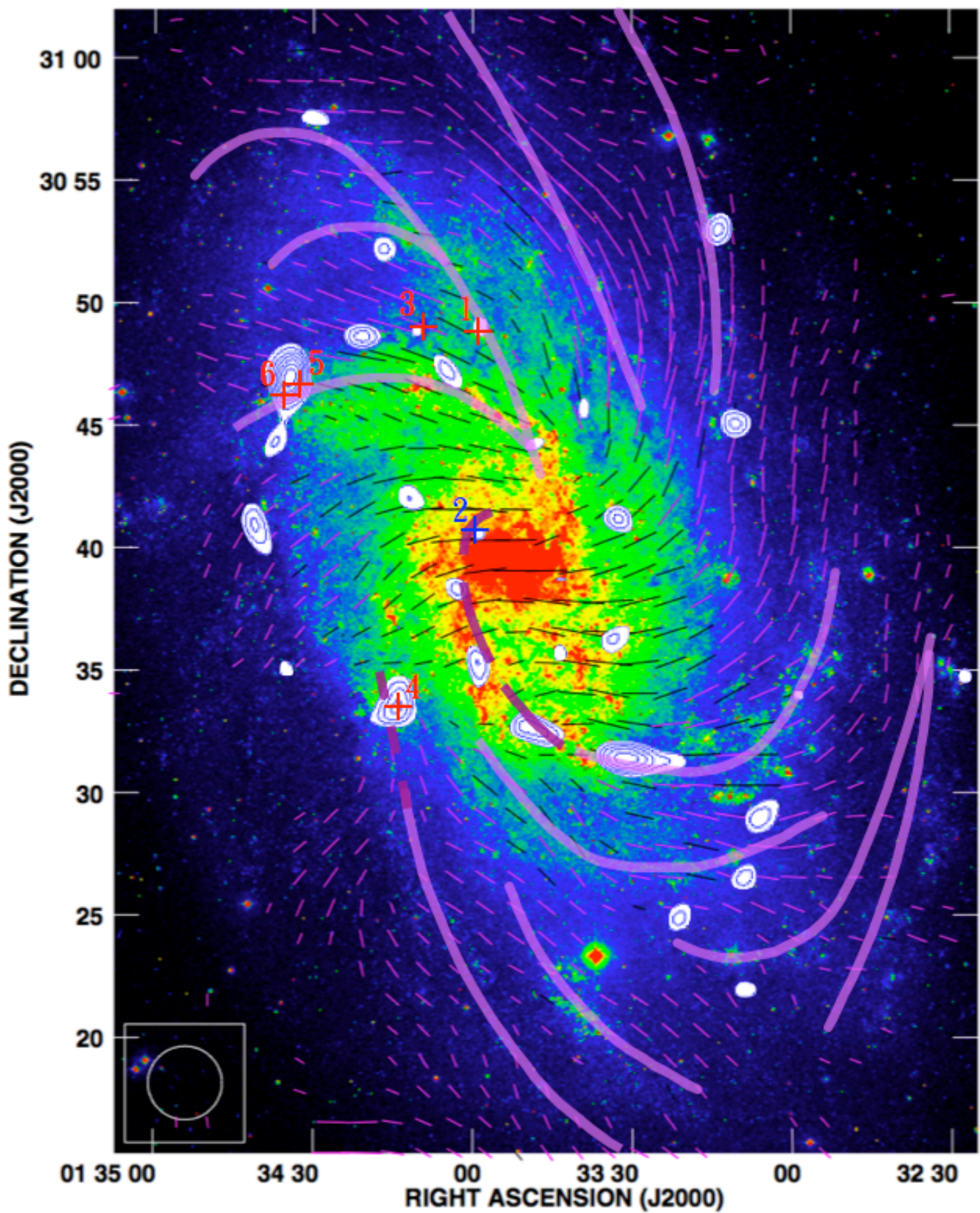

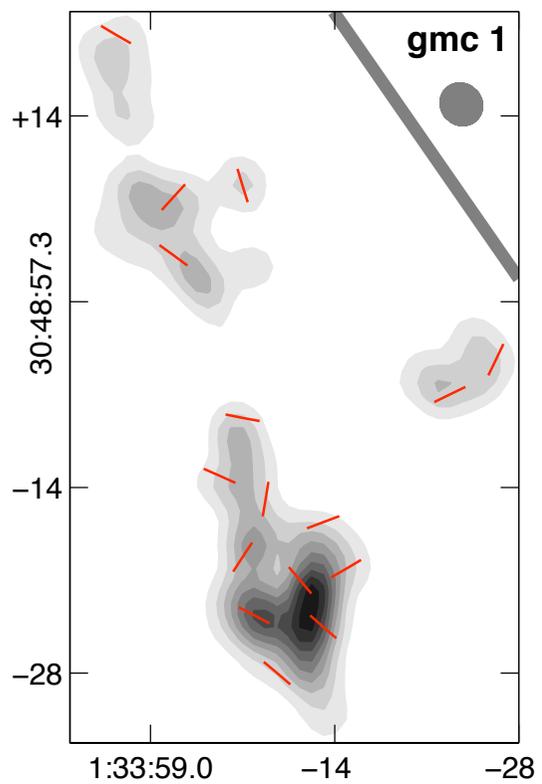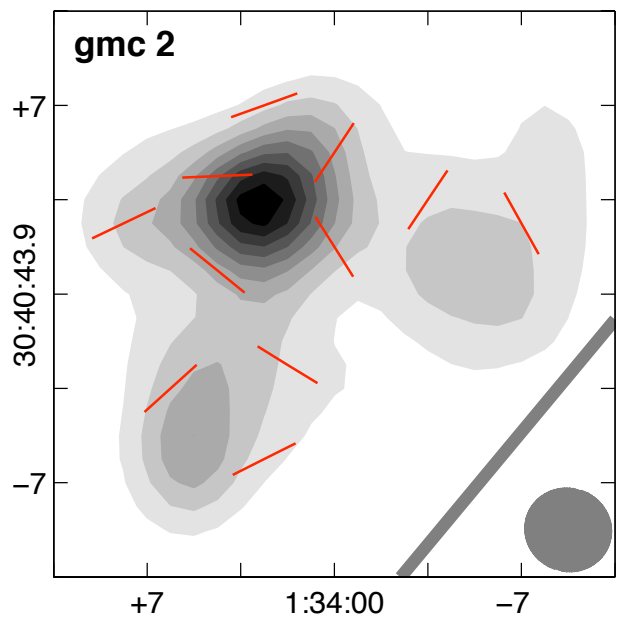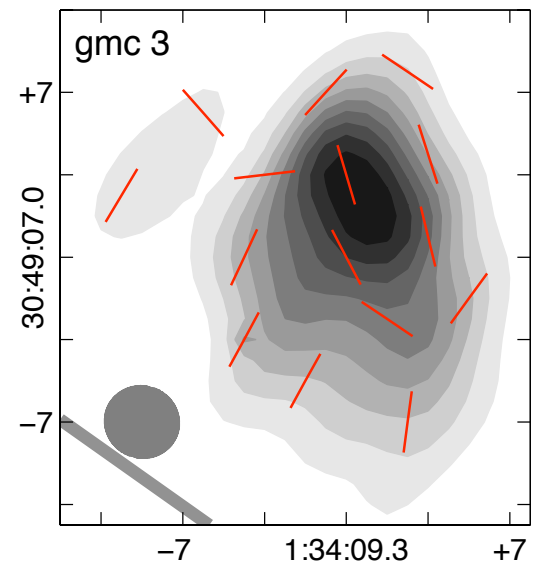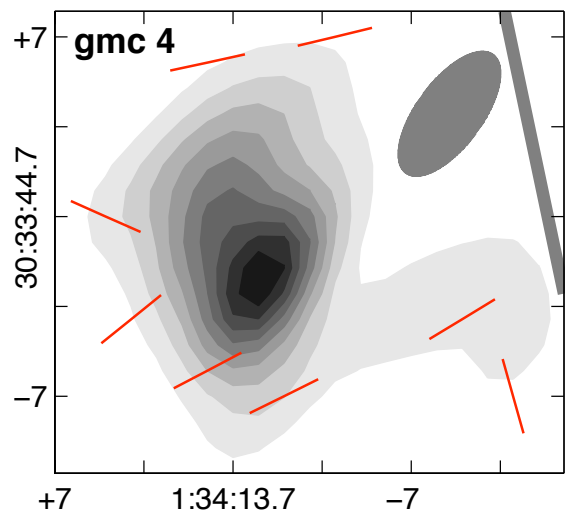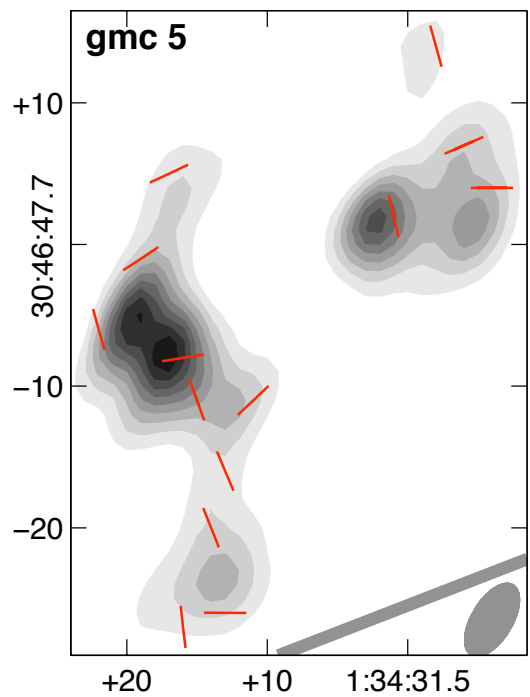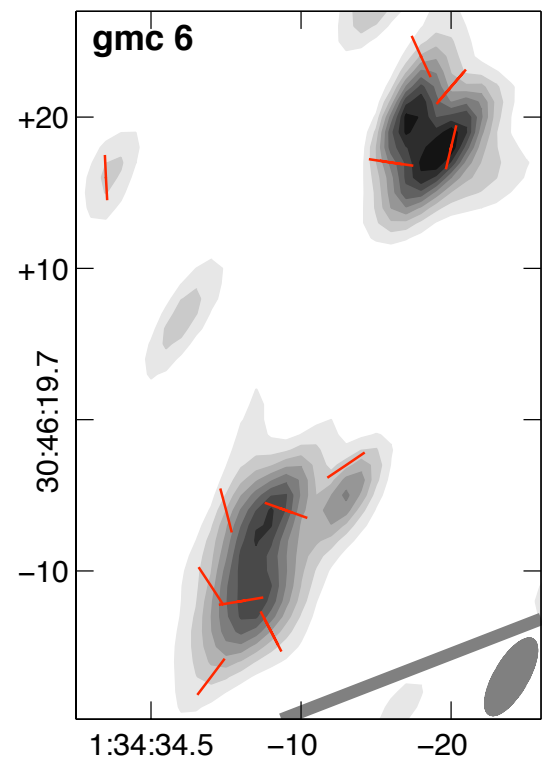

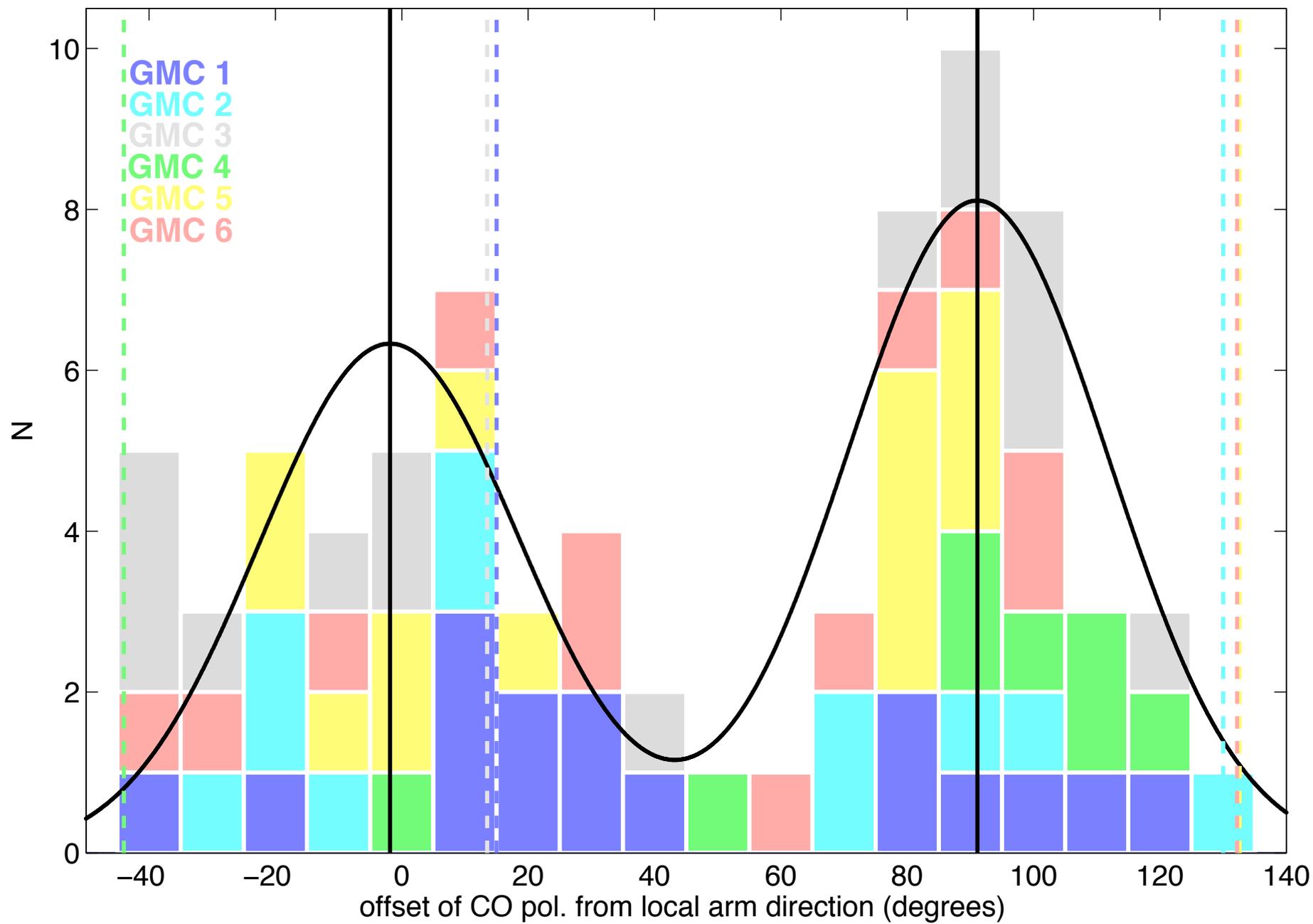

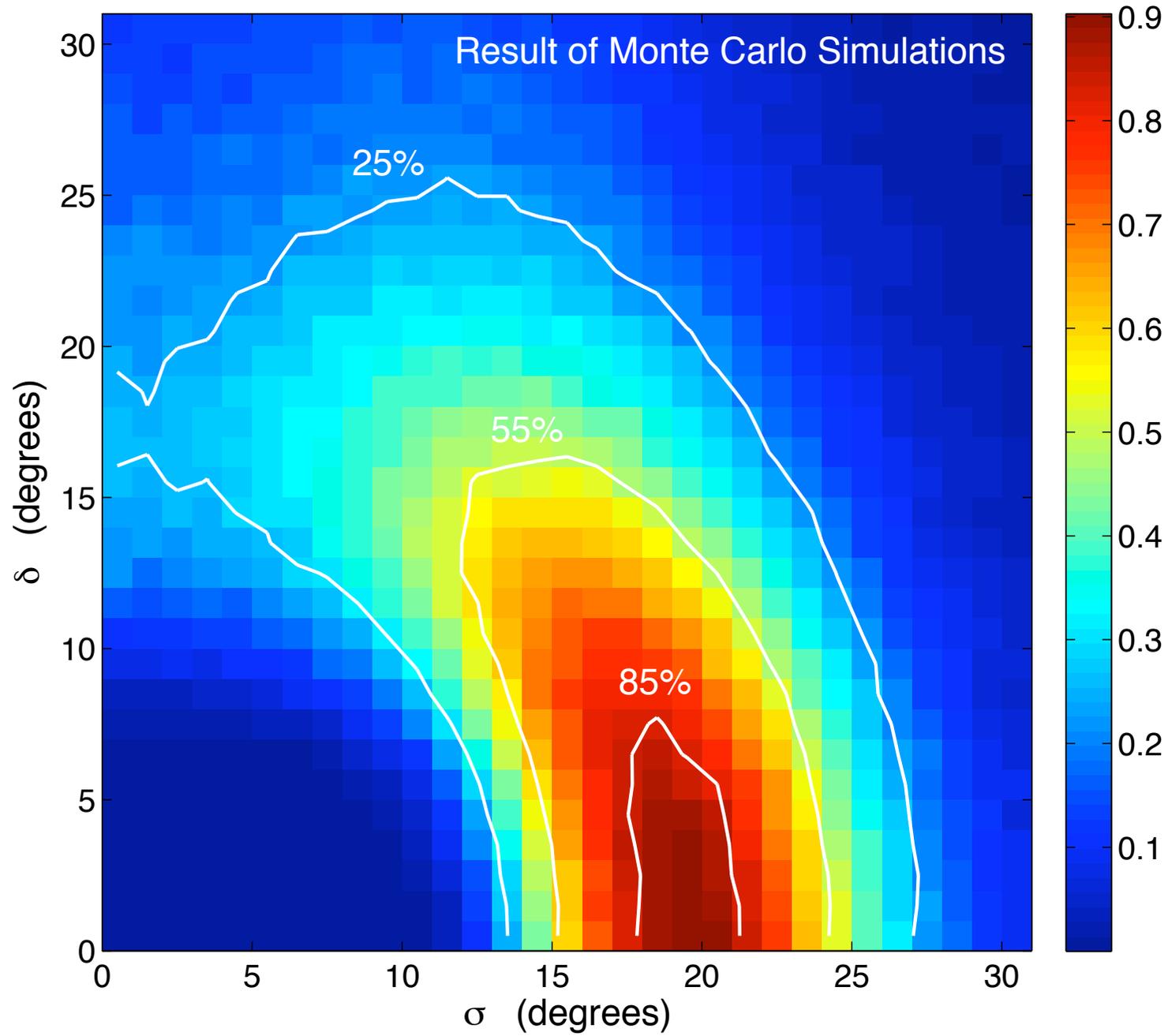